\documentclass[letter,traditabstract]{aa}
\usepackage{graphicx}      
\usepackage{txfonts}
\usepackage{natbib}
\usepackage{epsfig}
\usepackage{subfigure}
\usepackage{url}
\bibpunct{(}{)}{;}{a}{}{,} 

\begin{document}

\title{Random mixtures of polycyclic aromatic hydrocarbon spectra match interstellar infrared emission}

\author{Marissa J.~F.~ Rosenberg \inst{1} \and Olivier Bern\'e
  \inst{2,3} \and C. Boersma \inst{4}}

\institute{Leiden Observatory, Leiden University, P.O. Box 9513, NL- 2300 RA Leiden, The Netherlands \\\email{rosenberg@strw.leidenuniv.nl} \and 
  Universit\'e de Toulouse; UPS-OMP; IRAP; Toulouse, France \and CNRS; IRAP; 9 Av. colonel Roche, BP 44346, F-31028 Toulouse cedex 4, France \and
  NASA Ames Research Center, MS~245-6, Moffett Field, CA 94035-0001, USA}

\date{}

\abstract{The mid-infrared (IR; 5-15~$\mu$m) spectrum of a wide variety
  of astronomical objects exhibits a set of broad emission features at 6.2, 7.7, 8.6, 11.3 and 12.7 $\mu$m.
  About 30 years ago it was proposed that these signatures are due to
  emission from a family of UV heated nanometer-sized carbonaceous molecules
  known as \emph{polycyclic aromatic hydrocarbons} (PAHs), causing
  them to be referred to as aromatic IR bands (AIBs). Today, the
  acceptance of the PAH model is far from settled, as the
  identification of a single PAH in space has not yet been successful
  and physically relevant theoretical models involving ``true'' PAH
  cross sections do not reproduce the AIBs in detail.  In this paper,
  we use the NASA Ames PAH IR Spectroscopic Database, which contains over 500 quantum-computed spectra, in conjunction with a
  simple emission model, to show that the spectrum produced by any
  random mixture of at least 30 PAHs converges to the same
  \emph{'kernel'-spectrum}. This kernel-spectrum captures the essence
  of the PAH emission spectrum and is highly correlated with
  observations of AIBs, strongly supporting PAHs as their source.
  Also, the fact that a large number of molecules are required implies
  that spectroscopic signatures of the individual PAHs contributing to
  the AIBs spanning the visible, near-infrared, and far infrared
  spectral regions are weak, explaining why they have not yet been
  detected. An improved effort, joining laboratory, theoretical, and
  observational studies of the PAH emission process, will support the
  use of PAH features as a probe of physical and chemical conditions
  in the nearby and distant Universe.} 


\titlerunning{Variations of PAH Features}

\authorrunning{Rosenberg, M.~J.~F. et al.}

\maketitle

\section{Introduction}

The mid-infrared spectrum (IR; 5-15 $\mu$m) of astronomical objects is dominated by band emission 
strongest at 6.2, 7.7, 8.6, 11.3 and 12.7 $\mu$m and were referred to as the unidentified IR (UIR) bands. 
Discovered during the 1970's, this family of emission features are now generally
attributed to mixtures of polycyclic aromatic hydrocarbons; large,
chicken-wire shaped molecules of fused aromatic rings, and closely
related species \citep[e.g.,][and references
therein]{2007ApJ...657..810D, tie08,2014A&A...561A..82S}. In space, free floating PAH molecules become
highly vibrationally excited upon the absorption of a single photon
(ultraviolet~\textendash~near-infrared; \citealt{2005ApJ...629.1183M,2007ApJ...657..810D}). Relaxation occurs through emission of IR photons at characteristic wavelengths, leaving the
tell-tale spectroscopic fingerprints of aromatic molecules. The
features that comprise this apparently universal spectral signature
dominates the mid-IR spectrum of many bright astronomical objects and
contain a wealth of information about the physical conditions in the
emitting regions \citep[e.g.,][and references
therein]{job11}. However PAHs are no silent witnesses,
they are active participants and drive many aspects of the galactic
evolution from cold collapsing clouds to stellar death (see \citet{tie08}).

Initially, the AIB spectrum observed in the ISM of dusty galaxies
seemed invariant, yet as observations improved variations in relative
band strengths and subtle shifts in peak positions and profiles were
reported. In particular, an evolution of the ratio of the short (i.e.,
at 6.2, 7.7 and 8.6~$\mu$m) to long wavelength bands (i.e., at 11.2
and 12.7~$\mu$m) was identified, both in the galactic \citep{bre89,
  job96} and extragalactic \citep{gal08} ISM. These variations have
been attributed to a varying degree of PAH ionization \citep{szc93,
   hud95b,hud95a}.

However, the acceptance of the PAH model is far from settled.  Recently, the PAH hypothesis was contested by \citet{kwo11} who propose that this emission arises from complex organic solids with disorganized structures. These claims
were quickly rebutted \citep{2012ApJ...760L..35L}. Nonetheless, the
lack of the identification of a single PAH molecule in space remains
troublesome for some.

In this Letter, we show how the astronomical AIB spectrum can be
reproduced from first principles using the quantum-chemically
calculated PAH spectra from the NASA Ames PAH IR Spectroscopic
Database (\url{www.astrochem.org/pahdb}). More specifically, we show
how spectroscopic models are 1) able to explain the observed
universality, and 2) reproduce the spectrum. Furthermore, we iterate
on the fact that the identification of a single PAH will likely remain
illusive.

This work is presented as follows. In Section~2, the AIB spectra from
a range of different physical environments are presented. Then in
Section~3, using the quantum-chemically calculated PAH spectra,
astronomical AIB spectra are re-created by mixing the PAHs in the
database with random abundances. The minimum number of PAHs required
to reproduce observed spectra is established in Section~4 and
Section~5 states our conclusions. The appendix addresses database biases, 
details the statistical analyses and goes into blind signal separation techniques.

\section{Observed PAH spectrum}
\begin{figure}
  \resizebox{\hsize}{!}{\includegraphics{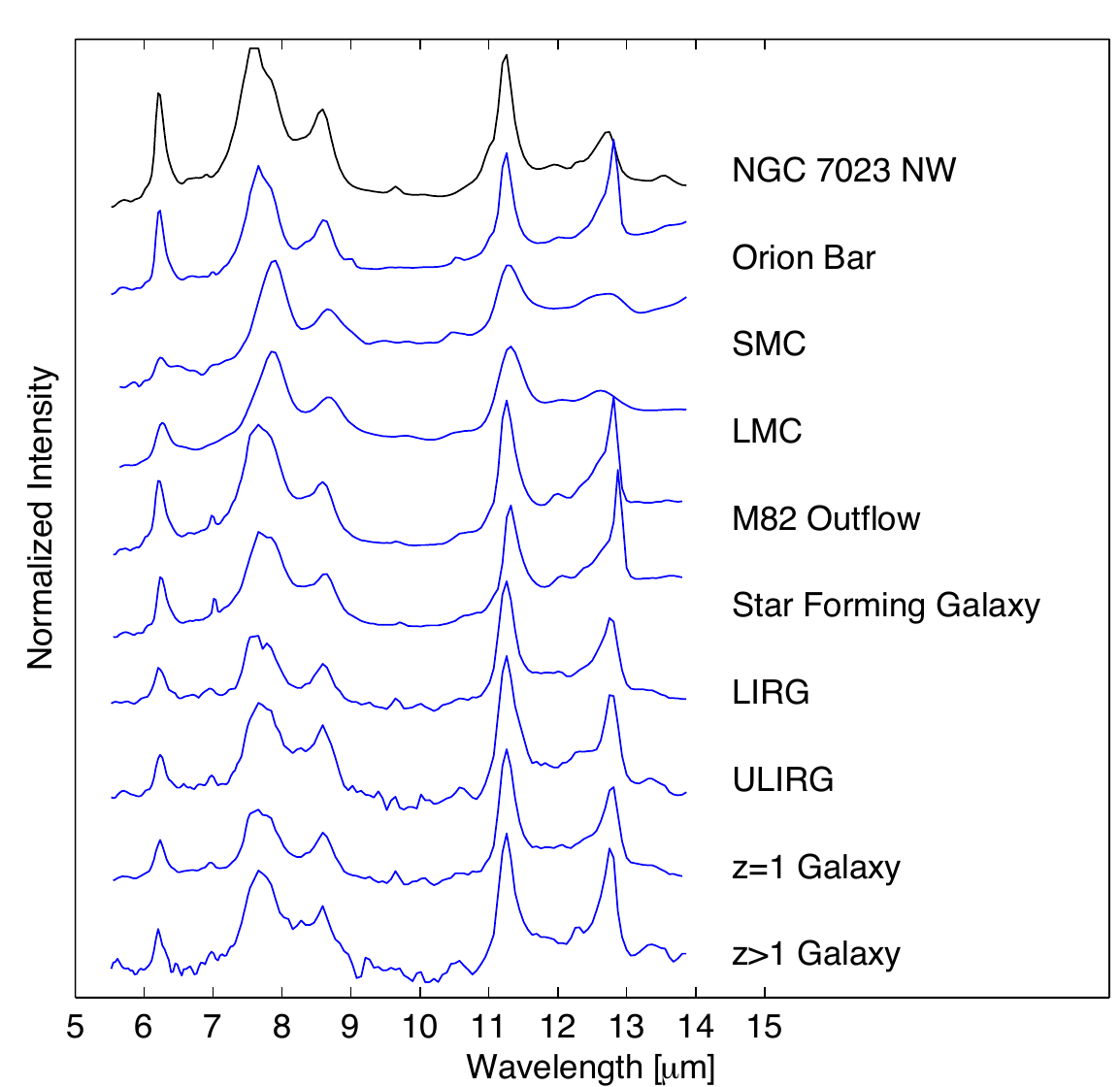}}
  \caption{Mid-infrared spectra of the interstellar medium in regions of the Milky Way and other galaxies obtained with the IRS instrument onboard the Spitzer Space Telescope. The Orion Bar is a UV irradiated molecular ridge in the Orion star-forming region. SMC and LMC refer to the Small and Large Magellanic Clouds. M82 is a nearby galaxy with intense star formation activity. The Luminous and Ultraluminous Infrared Galaxies (respectively LIRG and ULIRG) spectra are templates obtained by stacking the spectra from a large sample of galaxies \citep{das09}. The $0<z<1$ and $1<z<2$ spectra, where z means redshift, were obtained by stacking the spectra from a large sample of galaxies \citep{das09} and are presented in their rest frame.  These spectra represent galaxies in the early universe.}
  \label{fig:obs}
\end{figure}

Between 2003 and 2009, NASA's Spitzer Space Telescope \citep{wer04}
observed numerous astronomical sources in the mid-infrared
(5-15~$\mu$m). These data have been made publicly available
and can be accessed
online\footnote{http://sha.ipac.caltech.edu/applications/Spitzer/SHA/}. We
have compiled a sample of the mid-IR spectra.  These
data were obtained with Spitzer's infrared spectrograph (IRS; \citealt{2004ApJS..154...18H}),
utilizing its short wavelength, low resolution module (SL), and are
shown in Figure~\ref{fig:obs}. These spectra include observations of
the interstellar medium (ISM) ranging from specific regions in our
galaxy to galaxies up to redshifts $z>1$.

The spectra in Figure~1 span a wide range of objects, physical
conditions, star formation activity/history, age, metallically, and
mass. While showing small differences in detail, regardless of their
environment, all spectra resemble each other in peak position and
profile and are dominated by emission features
centered around 6.2, 7.7, 8.6, 11.2 and 12.7~$\mu$m. The similarity
between the spectra in Figure~\ref{fig:obs} is striking, with an
average correlation coefficient of 0.8. This high correlation between
observed spectra implies universal properties for the carrier(s) of
these AIBs.  In the following, we will attempt to reproduce these observations 
using quantum chemically calculated spectra and to explain the universal character of the observed AIB spectrum. 
 The black spectrum in Figure 1 is NGC 7023-NW and will be used as the typical PAH spectrum for comparison to the models in the remainder of the text.This is an environment with an intense UV radiation field, representative of star-forming regions, which dominate the mid-IR emission of galaxies.

\section{Results}

\subsection{NASA Ames PAH IR Spectroscopic Database}

Version 1.32 of the NASA Ames PAH IR Spectroscopic
Database is used, which contains 659 computed PAH spectra \citep{bau10,boe14}. From these,
548 were selected, excluding those spectra from PAHs containing
oxygen, magnesium, silicon and iron, as these are commonly not
considered to exist at high abundances, if at all, in space. The
selected set holds the spectra from a wide variety of PAHs; PAHs of
different sizes ($9\le N_{\rm C}\le384$), ionization states
(-,0,+,++,+++), and geometries. A discussion on
biases in the database can be found in Appendix~\ref{ap:bias}.

In order to compare astronomical AIB \emph{emission} spectra with
computed \emph{absorption} spectra from the NASA Ames PAH IR
Spectroscopic Database, an emission model is required. Here, the
\emph{single photon, single PAH} emission model is adopted, which is
fully implemented by the \textit{AmesPAHdbIDLSuite} \citep{bau10,
  boe14}. Each PAH is given an internal vibrational energy of 6.0~eV
and put through the entire emission-cooling cascade.

While the choice of emission profile and its width will not affect our
\emph{general} result, they do become important when making the direct
comparison to astronomical observations. Here Lorentzian emission
profiles are chosen with a fixed width of 15~cm$^{-1}$.  However,
anharmonicity will shift the peak position and broaden the bands. The
exact amount of this shift and broadening is dependent on the specific
molecule, temperature, and vibrational mode \citep{1992ApJ...401..269C, 1998ApJ...493..793C, 1995A&A...299..835J,
  2002A&A...388..639P, 2003ApJ...591..968O}. It is this complexity
that restricts the emission model from dealing with anharmonicity in
detail. However, an overall redshift of 15~cm$^{-1}$ is applied, which
is consistent with shifts for the out-of-plane bending modes of PAHs
at $\sim$900~K measured in the laboratory \citep{1995A&A...299..835J,
  2002A&A...388..639P}. The resulting emission
spectra are presented in the top panel of Figure~\ref{fig:mix} and
illustrate the diversity in spectral features between PAH species.

\subsection{Database Mixtures}
\label{sec:mix}
\begin{figure*}
\centering
\includegraphics[width=18cm]{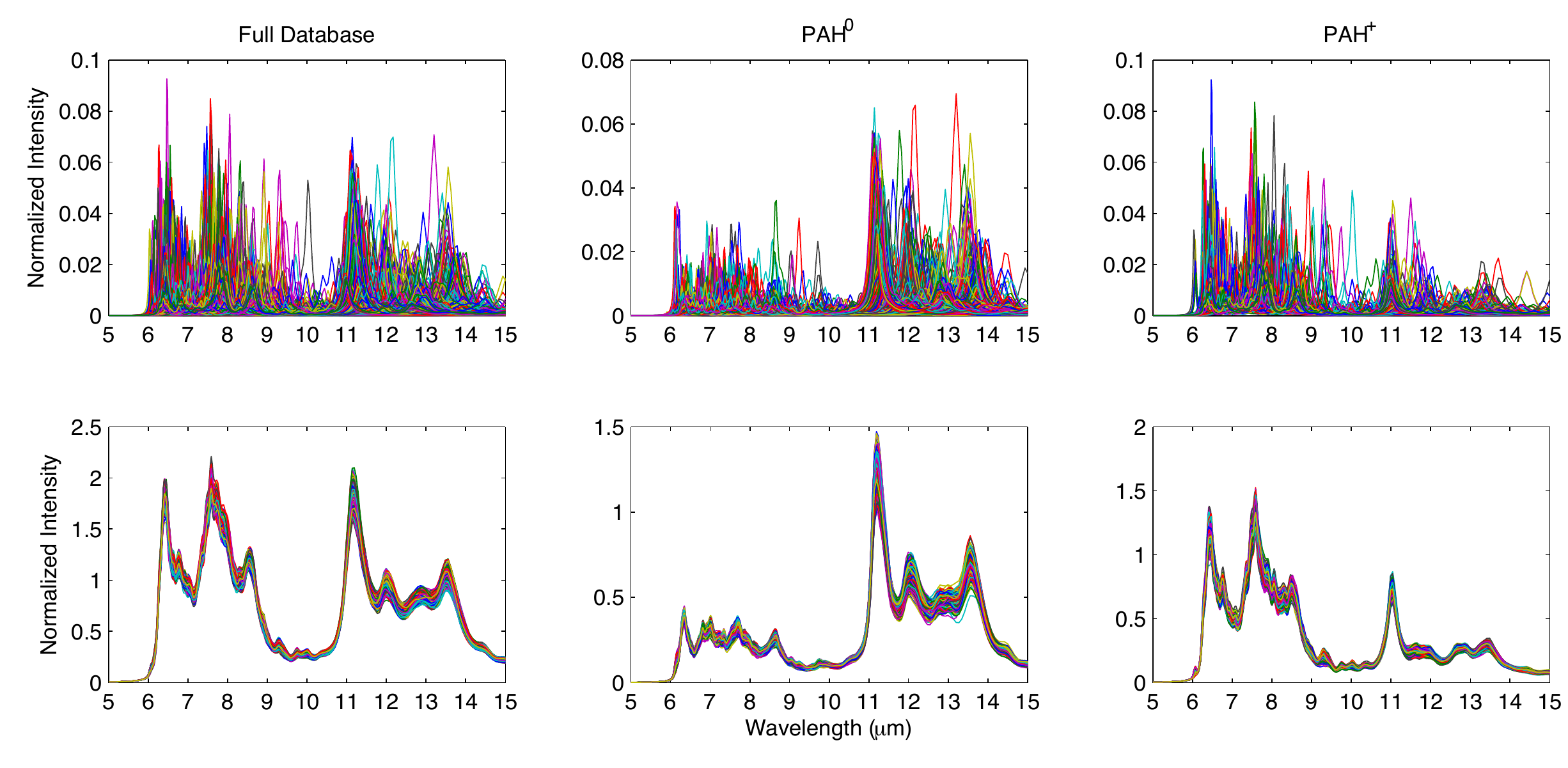}
  \caption{The effect of randomly mixing 5-15 $\mu$m PAH spectra \textbf{Left column:} Upper panel: Calculated emission spectra for all of the 548 pure or nitrogenated PAH species from the NASA Ames PAH IR Spectroscopic Database \citep{bau10, boe14}. Each color represents the spectrum of a different PAH species. The width of the PAH bands has been set to 15 cm$^{-1}$ and a redshift of 15 cm$^{-1}$ has been applied to mimic some effects of anharmonicity. These parameters have been selected to convert the database spectra from absorption into emission spectra and are close to values observed in the laboratory (Appendix~\ref{ap:bias}). Lower panel: 1000 random mixtures of the 548 different PAH emission spectra in the upper panel. Each color represents one of the 1000 mixtures, each mixture containing 548 species in different abundances. \textbf{Middle column:} Same as left column but considering only PAH cations (PAH$^+$). \textbf{Right column:} Same as left column but considering only PAH neutrals (PAH$^0$).}
  \label{fig:mix}
\end{figure*}

From the 548 PAH emission spectra calculated in Section 3.1, 1000 random mixtures were created
by assigning each 5-15~$\mu$m PAH spectrum a random
abundance between 0 and 1 and then averaging.
The results are presented in the left panel of
Figure~\ref{fig:mix}. The similarity between the 1000 mixtures is
astonishing, with an average correlation coefficient of 0.96 (see Appendix~\ref{ap:pdf}).

We can continue this analysis for the different charge states of the
PAH species. From the 254 and 222 PAH cations and neutrals,
respectively, we create 1000 random mixtures of each. The middle and right
panels of Figure~\ref{fig:mix} present these mixtures for the PAH
cations and neutrals, respectively. Consistent with the results from
the entire dataset, the mixed spectra are extremely similar. Yet
unlike the full dataset, clearcut difference appear between the two
separate charge sets. These differences between neutral and cations
are in agreement with those mentioned earlier \citep{szc93, hud95a,
  hud95b, all99}.  We refer to the average spectrum from the 1000 mixtures
calculated above as ``kernel'' spectra.

\section{Analysis}

\subsection{Comparison with Observations}

\begin{figure*}
  \centering
  \includegraphics[width=18cm]{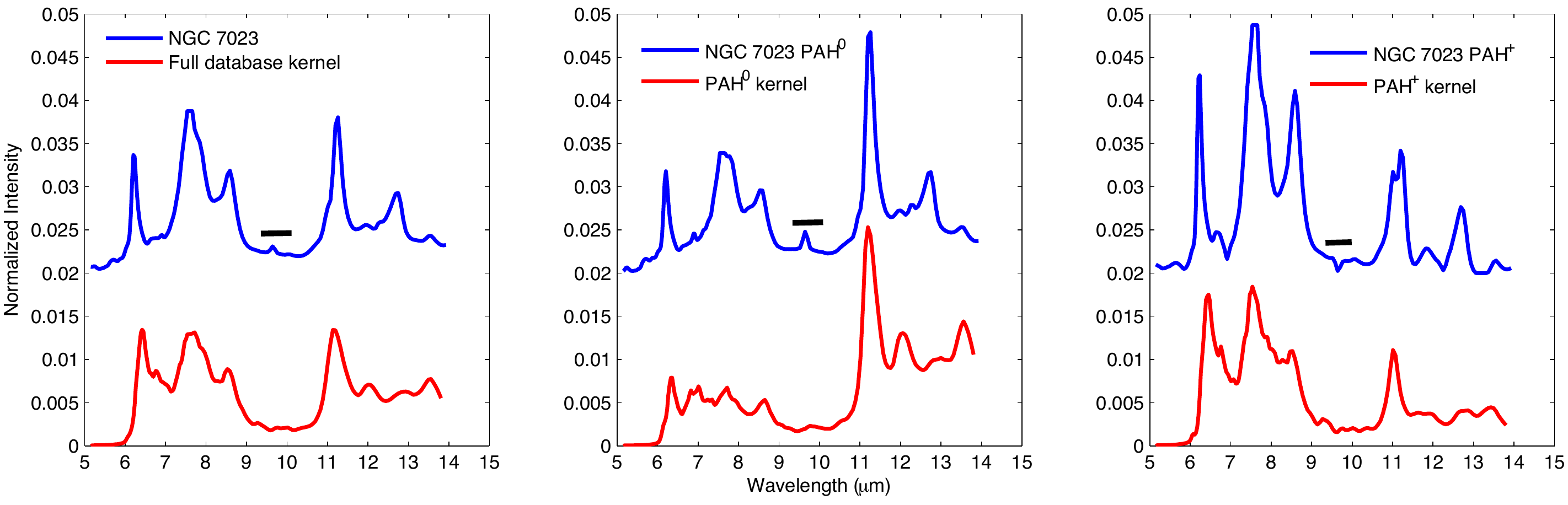}
  \caption{Comparison of the 5-15~$\mu$m kernel spectra (Figure\ref{fig:mix}) with the observations and components from a blind signal decomposition (BSS) of the reflection nebula NGC 7023. {\bf Left Panel:} Comparison of the spectrum of NGC~7023 (blue) and the kernel spectrum resulting from the 1000 random mixtures of 548 computed PAH spectra. {\bf Middle Panel:} Comparison between the astronomical neutral PAHs (PAH$^0$) spectrum extracted from the observations of NGC~7023 (blue) using blind signal separation techniques (\citealt{rap05, ber07, ber10};Appendix~\ref{ap:bss}) and the kernel spectrum (red) resulting from random mixtures of computed neutral PAH spectra. {\bf Right Panel:} Same as the middle panel but now for cation PAH species (PAH$^+$). In all panels, solid black line indicates the pure rotational line of the molecular hydrogen molecule at 9.7~$\mu$m.}
  \label{fig:comp}
\end{figure*}

In the left panel of Figure~\ref{fig:comp}, we compare the kernel spectra directly to
astronomical observations of the reflection nebula NGC~7023;
representative of the galactic ISM. The kernel spectrum of the full
database matches the salient characteristics of the observed NGC~7023
spectrum well. Some of the discrepancies between the kernel spectra
and observations can be attributed to the employed emission model,
which is unable to treat anharmonicity \citep{bas11}.  In addition, we
do not reproduce a strong 12.7 $\mu$m feature, which is observed in
most PAH spectra.  Recent work attributes this feature to long chains of
aliphatic PAHs (Candian et al. in prep.), which are under-represented in the
database.  Similarly, the 6.2 $\mu$m feature has been attributed to curved PAHs \citep{C4SC00890A}, which are not yet represented in the database, and thus we cannot reproduce this feature.

Figures.~\ref{fig:comp}b)~\&~\ref{fig:comp}c) compare the kernel
spectra for PAH cations and neutrals, respectively, with the
corresponding spectra extracted from NGC~7023 separated using blind signal
separation techniques \citep{rap05, ber07, ber10}
(Appendix~\ref{ap:bss}). The agreement between the PAH kernel spectra
and the blind signal separations components extracted from the
observations, is striking for both neutral and cationic PAHs. We find a correlation coefficient of 0.74, comparing the 5-15 $\mu$m regions, 
and a coefficient of 0.94 if we exclude the 6.2 and 12.7 $\mu$m features.

\subsection{Statistical Analysis}
In order to determine the minimum number of PAH species necessary to reproduce the full database kernel spectrum with spread comparable to observations, we conduct a statistical analysis. The details of this analysis can be found in Appendix C. The general procedure involves making many mixtures (as in Section 3.2), but varying the number of PAH species included in the mixtures, from 10-100. The analysis shows that by increasing the number of species in the sample, the resulting variations between the spectra decrease. It can be shown that with $\sim$ 30 or more PAHs in the mixture, the variations in the modeled spectra are within the amount observed int the AIBs. This analysis relies on a biased sample (database and observations) and an arbitrary norm, therefore the number 30 should not be considered as a sharp limit (see appendix C for details). Nonetheless, these results indicate, that a large number of PAH molecules can reproduce the AIBs.

\section{Discussion}

The results presented here show that any random mixture of many (more
than $\sim$30) PAH molecules, given the limitations of the adopted emission
model, can reproduce the salient spectral characteristics of the
astronomical AIB spectrum to a remarkable degree. In this respect, the
similarity between the AIB spectra observed across the Universe can be
understood as being due to the presence of a sufficient number of
varying interstellar PAHs (on the order of 30), which, when emitting
together, produce the Universal 'kernel' spectrum. While this spectrum
can only arise from a mixture containing a sufficient number of
varying PAH species, given the limitations of the observational
spectra and databases, it does not seem to require the presence of a
specific PAH. Instead, the observed variations can be attributed to
changes in the overall, global, characteristics of the PAH population
such as charge, structure, and so on. 

Additionally, this implies it will be challenging to identify a
particular interstellar PAH by any other spectroscopic means, in
particular by electronic transitions in the UV to near infrared range,
far-IR vibrational bands, or pure-rotational lines at millimeter
wavelengths (for PAHs with a dipole moment) as spectral signatures are
damped by a factor that is -- to first order -- proportional to the
number of different species. This is a large part of the reason why
attempts to identify these signatures have not succeeded so far
\citep{2006A&A...456..161M,pil09,2011A&A...530A..26G}. To make
progress in these other wavelength regions, where specific PAH
identification is possible, will require much higher signal-to-noise
observations coupled with very high precision laboratory spectroscopy.

Recently, the idea that PAH could be the carriers of the AIBs was contested by \citet{kwo11} who propose that, instead, these same bands arise from complex organic solids with disorganized structures. This latter work, based on an analysis of archival infrared spectroscopic data, relies on an empirical decomposition of the observed spectra and does not include theoretical emission cross sections nor any modeling of the emission mechanism for the proposed carriers. Our analysis instead shows that random mixtures of many ($>30$) theoretical emission spectra of PAH molecules correlate with observations to a high degree. In addition, with our model, we are able to explain the remarkable consistency of the shape of the PAH spectrum in the universe as well as the subtle spectral variations observed in the Milky Way. On this solid basis, we conclude that PAHs are better candidates for the AIBs than the structures proposed by \citet{kwo11}.

Finally, as properties such as ionization are tangible and closely linked to local conditions (i.e., UV-field, and electron density and the temperature of the gas), improved knowledge of PAHs and their emission mechanism will improve the use of AIBs as probes of local and distant astrophysical environments. Future instruments, such as the mid-infrared instrument (MIRI) on the James Webb Space Telescope (JWST), and the Mid-infared E-ELT Imager and Spectrograph (METIS) on the European Extremely Large Telescope (E-ELT) will open up the wealth of information hidden in the details of the AIB spectra. Although these instruments will not be able to identify specific PAH species due to the band overlap, they will be able to trace in finer details the general properties of the PAH family and to relate these properties to local physical conditions.

\bibliographystyle{aa}
\bibliography{full_bib.bib}

\appendix
\section{Database Biases}
\label{ap:bias}

The content of the computed part of the NASA Ames PAH IR Spectroscopic
Database has some intrinsic biases \citep{bau10, boe14}. These biases
originate historically from a limit in available computational power,
that smaller PAH species (N$_C<20$) are more easily calculated, and a
focus on astrophysically relevant species, i.e., 'pure',
cata-condensed, neutral and singly positively ionized PAH species
\citep{tie08}. The bias towards small PAHs in the database is somewhat
negated by the physics of the PAH emission process and the wavelength
range considered here (5-15 $\mu$m). Small PAHs get significantly
hotter than their larger counterparts upon absorbing the same photon
energy. This pushes more of the emission blue of the wavelength region
considered here. Similarly for large PAHs (N$_C>80$), which stay
significantly cooler, more of the emission is pushed red of the
wavelength region considered here. From a stability standpoint, the
family of cata-condensed, compact PAHs is very stable and hence more
likely to survive the rigors of interstellar space
\citep{all85}. These species are well represented in the database. The
database does undersample dehydrogenated PAHs, both in the levels of
dehydrogenation and the possible permutations. However, it has been
shown that the removal of only one or two hydrogens from
cata-condensed PAHs does not alter the spectrum much and that such
fully dehydrogenated species in space are probably rare
\citep{2013ApJ...776..102B}. Variations in peripheral hydrogen
adjacencies are reflected by variations in the 10-15 $\mu$m region of the
PAH spectrum \citep{hon2001}. As PAHs become increasingly larger,
while remaining compact, they obtain more and more straight
edges. This is reflected in their spectra by a strong 11.2 $\mu$m
feature. Adding more irregular PAHs to the database can alleviate some
of the current bias towards compact, straight edged PAHs in the
database. Of all studied hetero atom substitutions, nitrogen has shown
to be the most viable candidate, as its inclusion does not affect
PAHs stability, nitrogen is abundant in the circumstellar shells
around carbon rich AGB stars
\citep{all85,1989ApJS...71..733A,1989ApJ...341..372F}, the place where
PAHs are thought to be formed (\cite{2006A&A...447..213B}, and references therein) and their
known presence in meteorites \citep{1977GeCoA..41.1325H}. Other
substitutions either have little effect (e.g., silicon, magnesium) or
significantly disrupt the aromatic network, and therefore reduce the
stability of the PAH, e.g., oxygen \citep{2005ApJ...632..316H}. Singly
charged PAH anions are well represented in the database for the larger
PAH species. Considering detailed charge balance, doubly charged PAH
cations only become important in the more extreme astrophysical
environments and higher ionization states can safely be ignored. These
and other considerations regarding database biases and their
astrophysical relevance have also been discussed in
\citet{2013ApJ...769..117B}.  Our database mixed spectra are affected
by these biases.  The region in the spectrum most affected is the
10-15 $\mu$m range due to the underrepresentation of irregular PAHs.
This could also explain the disparity between the observations and
the database mixtures in these regions e.g., the weak 12.7 $\mu$m
feature.

\begin{figure}
  \resizebox{\hsize}{!}{\includegraphics[width=3cm]{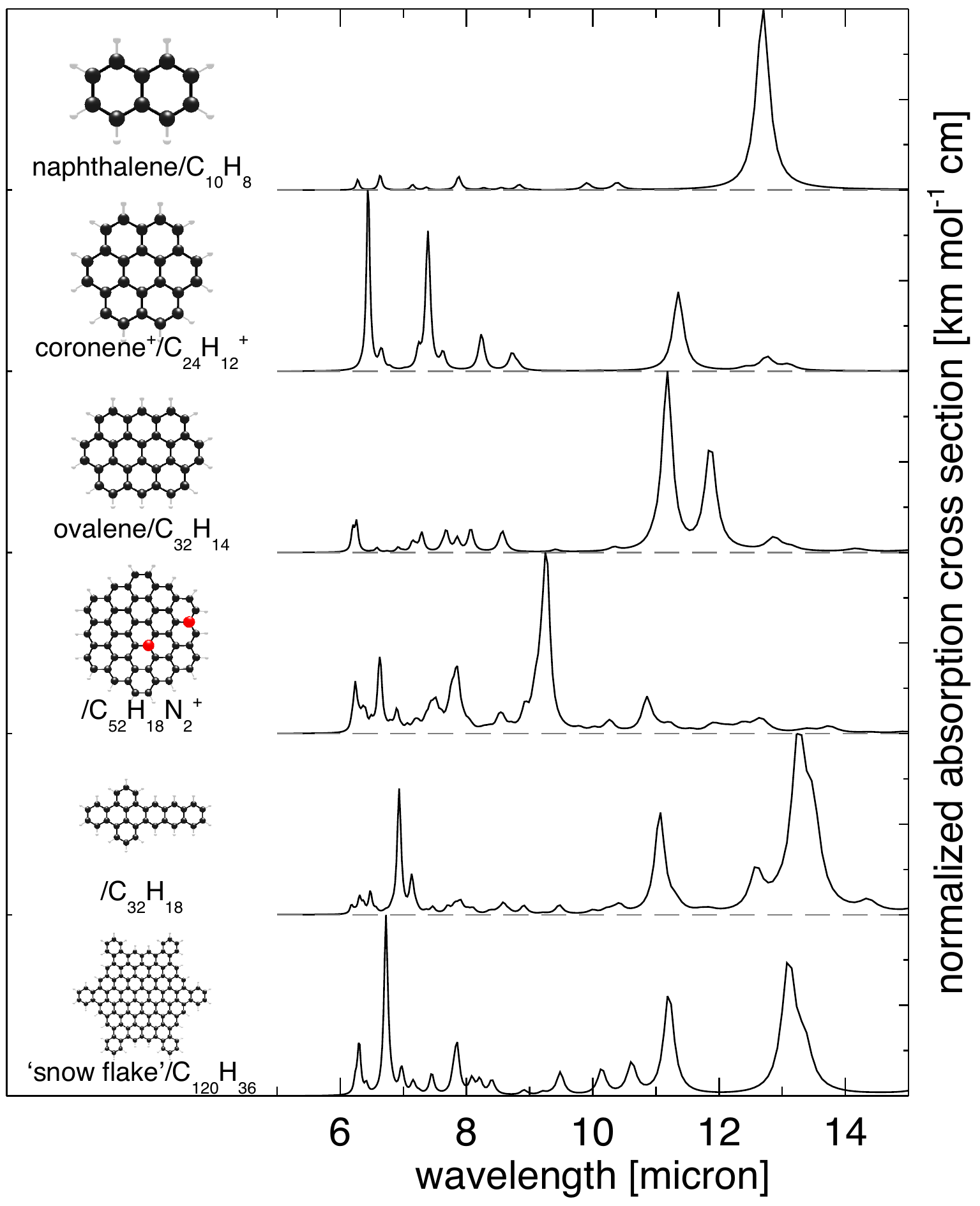}}
  \caption{Structure and DFT computed 5-15~$\mu$m vibrational infrared spectra of a selection of PAHs. PAHs are a class of carbonaceous molecules, which form a skeleton where carbon atoms are arranged in a honeycomb structure with hydrogen atoms sitting on the periphery. Additional atoms, such as Nitrogen, also can be present in the skeleton. For each species, the chemical formula, simple name (if it exists), and the corresponding mid-infrared spectrum calculated by DFT at 0~K are given. All data are taken from the NASA Ames PAH IR Spectroscopic Database \citep{bau10,boe14}.}
  \label{fig:db}
\end{figure}

\section{Correlation between the 1000 mixtures}
\label{ap:pdf}

Figure B1 shows the probability density function (PDF) of the correlation matrix of the 1000 mixtures (Figure 2 Left panel). The peak, average and median correlation coefficients are shown in blue, red and green, respectively. The 1-sigma variation around the mean is shown in yellow. The peak of the PDF (the most likely correlation) is found at 0.96 and the standard deviation is 0.023, i.e. 85\% of the correlations fall between 0.94 and 0.98. Only 4\% of the correlations are below 0.9.   The distribution is a sharp, narrow distribution, showing without a doubt that random PAH mixtures are indeed very alike.

\begin{figure}
  \resizebox{\hsize}{!}{ \includegraphics[width=3cm]{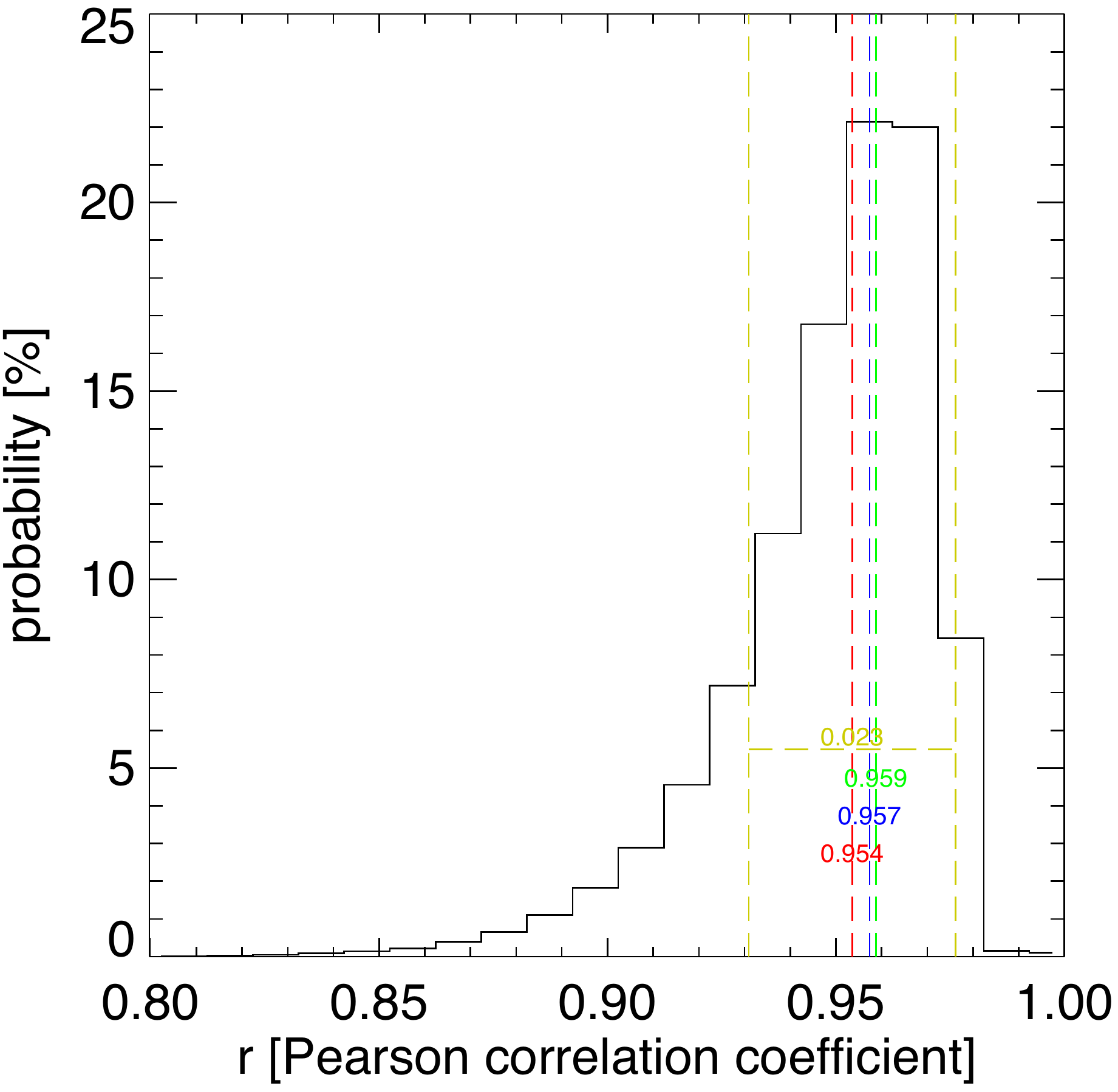}}
  \caption{Probability density function of the correlation coefficients between the average 5-15 $\mu$m spectra from 1000 mixtures of 548 species with random abundances between 0-1. The peak, average and median correlation coefficients are shown in blue, red and green, respectively. The 1-sigma variation around the mean is shown in yellow.}
  \label{fig:pdf}
\end{figure}

\section{Statistical Analysis Details}
\label{ap:stats}
First we concentrate on the database and apply the following procedure:  

\begin{enumerate}
  \item Randomly select $r$ spectra from the database.
  \item Create \emph{m} random linear combinations (mixtures) of the $r$ spectra by assigning each spectrum a random abundance between 0 and 1 such that; 
  \begin{equation}
    $X=AS$
  \end{equation}
  Where $X$ is an $m \times n$ matrix holding $m$ number of mixed
  spectra over $n$ wavelength bins, $A$ is an $m \times r$ abundance
  matrix, containing random numbers between 0 and 1, and $S$ is a
  $r\times n$ matrix holding the original set of database spectra.
  For this analysis we set $m$ to 100, creating 100 random mixtures
  each time.  
\item Repeat steps (1)-(2) $p$ times, randomly
  selecting a new set of $r$ PAH spectra each time (here we vary $r$
  between 10 and 100). Thus, $p$ spectra in matrix $X$ are created, which we will denote as $X_i$. We use $p$=100 for this analysis.  This means
  that there are $m$ random mixtures of $r$ spectra, and we re-select
  and re-mix those $r$ spectra $p$ times.
\item We find
  the maximum, minimum, and mean value of the spectra with index $i$, $X_i$, in the $X$ matrix
   and for every wavelength bin, $\lambda$. From
  these values we create three spectra, $S_{max}(\lambda)$,
  $S_{min}(\lambda)$, and $S(\lambda)$. The spectra
  $S_{max}(\lambda)$ and $S_{min}(\lambda)$ represent the boundaries
  within which any spectrum ($X_i$) falls. The mean spectrum $\bar{S(\lambda)}$
  is the kernel spectrum for that particular set of $m \times p$ mixed
  spectra.
\item We calculate the Euclidian distance from the minimum
  and maximum spectra with respect to the mean:
  \begin{equation}
    N_{min}=\sqrt{\sum_{\lambda=1..n}(S_{min}-\bar{S})^2}
  \end{equation}
  \begin{equation}
    N_{max}=\sqrt{\sum_{\lambda=1..n}(S_{max}-\bar{S})^2}
  \end{equation}
  and we define $N_r$ as
  \begin{equation}
    N_{r}= N_{min}+N_{max}
  \end{equation}
  which is a measurement of the maximum variation in the mixtures for
  each set of $r$ spectra. 
  \end{enumerate}

Now we consider the observations and calculate $N_{obs}$ using the
following steps:

\begin{enumerate}
\item Subtract a linear baseline (corresponding to the emission from
  another dust component) for each observed spectrum
\item Out of the 10 observed rest-frame spectra, define a minimum,
  maximum, and mean at each wavelength bin.
\item Similarly to step 4 in the database analysis, we find
  the maximum, minimum, and mean value of the \emph{observed} spectra with 
  index $i$, $X_i$, in the $X$ matrix
   and for every wavelength bin, $\lambda$. From
  these values we create three spectra, $S_{obs,max}(\lambda)$,
  $S_{obs,min}(\lambda)$, and $S_{obs}(\lambda)$. The spectra
  $S_{obs,max}(\lambda)$ and $S_{obs,min}(\lambda)$ represent the boundaries
  within which any of the observed spectra ($X_i$) fall. The mean spectrum $S_{obs}(\lambda)$
  is the average of the particular set of rest-frame observations presented in Figure~\ref{fig:obs}.

 Similarly to (6)
  for the database spectra, we define:
  \begin{equation}
    N_{obs,min}=\sqrt{\sum_{n}(S_{obs,min}-\bar{S}_{obs})^2}
  \end{equation}
  \begin{equation}
    N_{obs,max}=\sqrt{\sum_{n}(S_{obs,max}-\bar{S}_{obs})^2}
  \end{equation}
  \begin{equation}
    N_{obs}= N_{obs,min}+N_{obs,max}
  \end{equation}
  \end{enumerate}

\subsection{Comparison between database and observations statistics}
  
\begin{figure}
  \resizebox{\hsize}{!}{\includegraphics{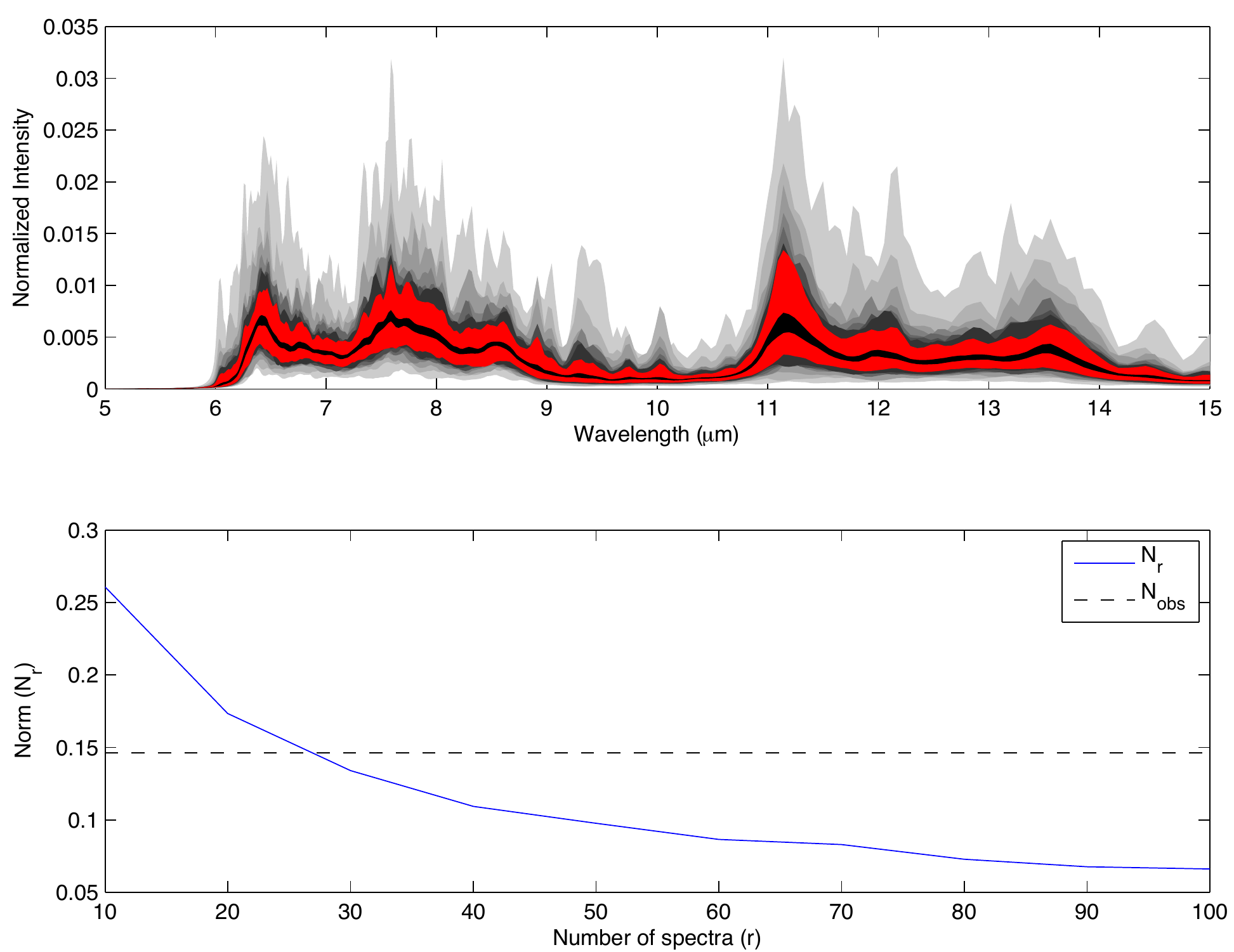}}
  \caption{Top panel: range of variations in the kernel spectra as a function of the number of PAH species considered in the mixture: 10 species (light grey) through 90 species (dark grey). In red is the range for 100 species and in black the range for 548 species. Bottom panel: evolution of the norm $N_r$ which captures the variations in the kernel spectra (blue line) and the norm $N_{obs}$ which captures the variations in the observations (see text for details).}
  \label{fig:dist}
\end{figure}

  In Figure~\ref{fig:dist}, we present the results of the statistical analysis
  in a graphical way. The top panel of Figure~\ref{fig:dist} shows the shaded regions
between $S_{min}$ and $S_{max}$, which highlights the boundaries of
the $m\times i$ mixtures of $r$ spectra.  The lightest grey shaded
region represents $r=10$, and increasingly darker greys represent
$r=20-90$.  The red region is where $r=100$ and the black region is
when $r=548$, the whole database.  It appears clearly that, by increasing
the number of species in the sample, the resulting variations between the kernel
spectra decrease. This can be investigated more quantitatively, by
following the evolution of the norm $N_r$ as a function of the number
of species present in the mixture $r$. This is done in the bottom panel
of Figure~\ref{fig:dist} where the decrease of $N_r$ can be seen clearly. 
One way to compare the variations of the observed 
AIB spectrum with those present in the kernel spectra, is to compare
$N_r$ with $N_{obs}$ which has a constant value reported in 
Figure~\ref{fig:dist}. When $N_r< N_{obs}$, the spectral variations 
(in terms of Euclidian norm as defined in Appendix~\ref{ap:stats}) of the
 database mixture are within the spectral variations observed in PAH.
This happens when $r>30$ (Figure~\ref{fig:dist}). 

\section{Blind Signal Separation}
\label{ap:bss}

\emph{Blind Signal Separation} is commonly used to restore a set of unknown ``source" signals from
a set of observed signals which are mixtures, or combinations, of these original source signals, with unknown mixture parameters \cite{hyv01}.
Several methods and algorithms exist in in the literature. The astronomical PAH cation and neutral spectra presented in this Letter
were obtained with Lee and Seung's non-negative matrix factorization (\citealt{lee01}, NMF). NMF was applied to data of the reflection 
nebula NGC 7023 obtained with the Infrared Spectrograph onboard the Spitzer Space Telescope. Details on the procedure 
can be found in \citet{ber10}.

\listofobjects

\end{document}